\documentclass[square]{ws-procs975x65}
\usepackage{amsmath}
\usepackage{chapterbib}

\begin{document}
\title{Integrating out  Holographic QCD back to Hidden Local Symmetry\footnote{Talk given by S. M. 
at Nagoya Global COE workshop: ``Strong Coupling Gauge Theories in LHC Era''  in honor of
Toshihide Maskawa's 70th Birthday and 35th Anniversary of Dynamical Symmetry Breaking in SCGT (SCGT 09),
Dec. 8-11, 2009, Nagoya University, Nagoya, http://www.eken.phys.nagoya-u.ac.jp/scgt09/.
To be published by World Scientific Publishing Co., Singapore (eds. M. Harada, H. Fukaya, M.Tanabashi and K. Yamawaki).  }}

\author{$^{a}$Masayasu Harada, $^{b}$Shinya Matsuzaki, and $^{a}$Koichi Yamawaki}

\address{$^{a}$Department of Physics, Nagoya University,
                    Nagoya, 464-8602, Japan. \\ 
E-mail: harada@hken.phys.nagoya-u.ac.jp \\ 
E-mail: yamawaki@eken.phys.nagoya-u.ac.jp }

\address{$^{b}$Department of Physics,  
                    Pusan National University, Busan 609-735, Korea. \\ 
E-mail: synya@pusan.ac.kr}

\begin{abstract}
We develop a previously proposed gauge-invariant method to integrate out 
infinite towers of vector and axialvector mesons 
arising as Kaluza-Klein (KK) modes 
in a class of holographic models of QCD (HQCD). 
We demonstrate that HQCD  can be reduced to the chiral perturbation theory (ChPT) 
with the hidden local symmetry (HLS) (so-called HLS-ChPT) 
having only the lowest KK mode identified as the HLS gauge boson, 
and the Nambu-Goldstone bosons. 
The  ${\cal O } (p^4)$ terms in the HLS-ChPT  
are completely determined by integrating out 
infinite towers of vector/axialvector mesons in HQCD: 
Effects of higher KK modes are fully included in the 
coefficients. 
As an example, we apply our method to the Sakai-Sugimoto model. 
\end{abstract}

\bodymatter 

\section{Introduction}

Holography, based on gauge/gravity duality~\cite{Maldacena:1997re}, 
has been of late fashion to 
reveal a part of features in
strongly coupled gauge theories involving the application to 
QCD (so-called holographic QCD (HQCD)).  
There are two types of holographic approaches: One is called 
``top-down" approach starting with a stringy setting; 
the other is called ``bottom-up" approach beginning with a
five-dimensional gauge theory defined on an AdS (anti-de Sitter space) background. 
It is a key point to notice that in whichever approach
one eventually employs a five-dimensional gauge model with a
characteristic induced-metric 
and some boundary conditions on a brane configuration.

In the low-energy region, any model of HQCD is reduced to a certain effective hadron model in four-dimensions. 
Such effective models include vector and axialvector mesons 
as infinite towers of Kaluza-Klein (KK) modes 
together with the Nambu-Goldstone bosons (NGBs) 
associated with the spontaneous  chiral symmetry breaking. 
Green functions in QCD are evaluated straightforwardly 
from the effective model following the holographic dictionary. 
Full sets of infinite towers of exchanges of KK modes 
(vector and axialvector mesons) contribute to Green functions involving current correlators and 
form factors and could mimic ultraviolet behaviors in QCD, 
although such a hadronic description would not be reliable above a certain high-energy scale. 
This implies that appropriate/gauge-invariant holographic results require calculations including  
{\it full} set of KK towers, which
, however, would not be practical 
because of forms written in terms of an infinite sum.

It was pointed out~\cite{Hill:2000mu} that the infinite tower of KK modes is 
interpreted as a set of gauge bosons of the hidden local symmetries (HLSs)~\cite{Bando:1984ej,Harada:2003jx}. 
This implies an interesting possibility that, in the low-energy region,  
any holographic models can be reduced to the simplest HLS model, 
provided that the infinite tower of KK modes 
is integrated out keeping only the lowest one identified with the $\rho$ meson and its flavor partners. 
Effects from the higher KK modes would then be {\it fully} incorporated 
into higher derivative terms (${\cal O} (p^4)$ terms) in the HLS effective field theory as an extension 
of the conventional chiral perturbation theory (ChPT)~\cite{Gas:84}, 
so-called the HLS-ChPT~\cite{Tanabashi:1993sr,Harada:2003jx} 
which is manifestly gauge-invariant formulation and makes it possible to calculate 
any Green functions order by order in derivative expansion. 
Once holographic models are expressed in terms of 
the HLS-ChPT, one can even calculate meson-loop corrections 
of subleading order in $1/N_c$ expansion. This
would give a new insight into the HQCD which as it stands is valid only in the
large $N_c$ limit.
Indeed, in the previous work~\cite{Harada:2006di}, we proposed a consistent 
method of integrating out the 
infinite towers of vector and axialvector mesons 
in the Sakai-Sugimoto (SS) model~\cite{Sakai:2004cn,Sakai:2005yt} 
into the 
HLS-ChPT, with the ${\cal O} (p^4)$ terms explicitly given by the integrated higher modes
effects.  Then we were able to do the first  calculations of $1/N_c$ corrections to the SS model.

In this talk, we report the results of our work~\cite{HMY2} developing in detail 
the integrating-out method proposed in Ref.~\cite{Harada:2006di}. 
First of all, we work in a class of holographic models to introduce our integrating-out method. 
Next, as an example, we apply our procedure to the Sakai-Sugimoto (SS) model~\cite{Sakai:2004cn,Sakai:2005yt} to give the HLS model with a full set of ${\cal O}(p^4)$ terms determined.
 We then calculate the pion electromagnetic form factor to 
 demonstrate how powerful 
 our formulation is even before including the $1/N_c$ effects through loop. 
As we will see explicitly, the momentum-dependence of the form factor is evaluated 
including {\it full} set of contributions from KK modes without performing infinite sums. 
Our method can straightforwardly be applied to other types of holographic 
models such as those given in Refs.~\cite{Da Rold:2005zs,Erlich:2005qh}. 
More details 
are presented 
in Ref.~\cite{HMY2}.

\section{A gauge-invariant way to integrate out HQCD} 
\label{sec2}

In this section, 
starting with a class of HQCD models including  
the SS model~\cite{Sakai:2004cn,Sakai:2005yt}, 
we introduce a way to obtain a  low-energy effective 
model in four-dimension described only by the lightest vector meson identified as
$\rho$ meson based on the HLS together with 
the NGBs. 
Suppose that the fifth direction, spanned by the coordinate $z$,  
extends from  minus infinity to plus infinity ($-\infty < z < \infty$)~\footnote{ In an application to 
another type of  HQCD~\cite{Da Rold:2005zs}, the $z$ coordinate is 
  defined on a finite interval, which is different from the $z$ 
  coordinate used here. They are related by an appropriate
  coordinate transformation as done in 
  Refs.~\cite{Sakai:2004cn,Sakai:2005yt}. 
}. 
We employ a five-dimensional gauge theory which has a vectorial $U(N)$ gauge symmetry 
defined on a certain background associated with 
the gauge/gravity duality. 
As far as gauge-invariant sector such as the Dirac-Born-Infeld part of the SS model~\cite{Sakai:2004cn,Sakai:2005yt}  
is concerned, the five-dimensional action in large $N_c$ limit  can be 
written as~\footnote{  Models of HQCD having the left- and right-bulk fields 
such as $F_L, F_R$~\cite{Da Rold:2005zs,Erlich:2005qh} 
can be described by the same action as in Eq.(\ref{5d-action}) with a suitable $z$-coordinate transformation 
prescribed. } 
\begin{eqnarray}
S_5 = N_c   
        \int d^4x dz  
         \Bigg( 
               - \frac{1}{2} K_1(z) {\rm tr}[ 
                                                   F_{\mu\nu} F^{\mu\nu} 
                                                 ]
               +      K_2(z) M_{\rm KK}^2  {\rm tr }[ 
                                                   F_{\mu z} F^{\mu z} 
                                                  ] 
         \Bigg)
\,, \label{5d-action}
\end{eqnarray} 
where 
$K_{1,2}(z)$ denote a set of metric-functions of $z$ 
constrained by the gauge/gravity duality.  
$M_{\rm KK}$ is a typical mass scale of 
KK modes of the gauge field $A_M$ with $M=(\mu, z)$. 
The boundary condition of $A_M$ is 
chosen as 
$ 
A_M(x^{\mu},z = \pm \infty) = 0 
$. 
A transformation which does not change this boundary condition 
satisfies 
$\partial_M g(x^\mu, z)|_{z=\pm \infty}=0$, 
where $g(x^\mu, z)$ is the transformation matrix of the gauge symmetry. 
This implies an emergence of global chiral $U(N)_L \times U(N)_R$ symmetry in four-dimension 
characterized by the transformation matrices $g_{R, L}=g(z=\pm \infty)$.

Following Refs.~\cite{Sakai:2004cn,Sakai:2005yt,Harada:2006di}, 
we work in $A_z=0$ gauge. 
There still exists a four-dimensional gauge symmetry 
under which $A_\mu(x^\mu,z)$ transforms as 
$ A_\mu \to h \cdot A_\mu  \cdot h^\dagger - i \partial_\mu h \cdot h^\dagger $ 
with $h = h(x^\mu)$. 
This gauge symmetry is identified~\cite{Sakai:2004cn,Sakai:2005yt,Harada:2006di} 
with the HLS~\cite{Bando:1984ej,Harada:2003jx}. 
In the $A_z \equiv 0$ gauge, the NGB fields $\pi(x^\mu)$ disappear from 
the chiral field $U=e^{2i\pi/F_\pi}$ since $U\to 1$. 
They are instead included~\cite{Harada:2006di} in $A_\mu(x^\mu,z)$  
at the boundary as 
$  A_\mu|_{z=\pm \infty} = \alpha^{R,L}_\mu 
= i \xi_{R,L}\partial_\mu \xi_{R,L}^\dagger$, 
where $\xi_{L,R}$ form the chiral field $U$ as 
$U=\xi_L^\dag \cdot \xi_R$.  
Since $\xi_{L,R} \to h \cdot \xi_{L,R} \cdot g_{L,R}^\dag$~\cite{Bando:1984ej,Harada:2003jx} 
,  $\alpha_\mu^{R,L}$ transform as 
$ \alpha^{R,L}_\mu \to h \cdot   \alpha^{R,L}_\mu  \cdot h^\dagger  
- i \partial_\mu h \cdot h^\dagger $.

We introduce infinite towers of massive KK modes for vector $(V_\mu^{(n)}(x^\mu))$
and axialvector $(A_\mu^{(n)}(x^\mu))$ meson fields, 
where we treat $V_\mu^{(n)} $ as the HLS gauge fields 
transforming under the HLS as 
$ V_\mu^{(n)} \to h \cdot   V_\mu^{(n)}  \cdot h^\dagger  
- i \partial_\mu h \cdot h^\dagger $,  
while $A_\mu^{(n)} $ as matter fields 
transforming as $ A_\mu^{(n)} \to h \cdot A_\mu^{(n)} \cdot  h^\dag $.   
The five-dimensional gauge field $A_\mu(x^\mu, z)$ is now 
expanded as~\footnote{In Eq.(\ref{Amuxz}) 
we put a relative minus sign in front of 
the HLS gauge fields $V_\mu^{(n)}$ 
for a convention. }
\begin{equation}  
  A_\mu(x^\mu, z) 
  = \alpha_\mu^R(x^\mu) \phi^R(z) 
  +
  \alpha_\mu^L(x^\mu) \phi^L(z) 
 + 
\sum_{n=1}^\infty \left( 
A_\mu^{(n)}(x^\mu) \psi_{2n}(z)
-    
V_\mu^{(n)}(x^\mu) \psi_{2n-1}(z) 
\right)
  \,. \label{Amuxz} 
\end{equation} 
The functions $\{ \psi_{2n-1}(z) \}$ and $\{ \psi_{2n}(z) \}$ are the eigenfunctions 
satisfying the eigenvalue 
equation obtained from the action (\ref{5d-action}): 
$ -  K_1^{-1}(z) \partial_z (K_2(z) \partial_z \psi_n(z)) 
= \lambda_n \psi_n(z) $, 
where $\lambda_n$ denotes the $n$th eigenvalue. 
On the other hand, the gauge-invariance requires 
the functions $\phi^{R,L}(z)$ to be different from the eigenfunctions: 
  From the transformation properties for $A_\mu(x^\mu,z)$, $\alpha_\mu^{R,L}$, $A_\mu^{(n)}$, and $V_\mu^{(n)}$ 
we see that the functions $\phi^{R,L}(z)$, $\{ \psi_{2n-1}(z) \}$, and $\{ \psi_{2n}(z) \}$  
are constrained as 
\begin{equation} 
  \phi^R (z)+ \phi^L (z)- \sum_{n=1}^\infty \psi_{2n-1} (z) = 1 
\,. \label{cons:general}
\end{equation}
Using this, we may rewrite Eq.(\ref{Amuxz}) to obtain 
$  
A_\mu(x^\mu, z) 
 = \alpha_{\mu ||} (x^\mu) 
  + 
  \alpha_{\mu \perp}(x^\mu) (\phi^R (z) -\phi^L (z) ) + 
\sum_{n=1}^\infty 
A_\mu^{(n)}(x^\mu) \psi_{2n}(z)
+ 
\sum_{n=1}^\infty 
\left( \alpha_{\mu ||}(x^\mu) -    
V_\mu^{(n)}(x^\mu) 
\right)\psi_{2n-1}(z) $, 
where $  \alpha_{\mu ||,\perp} =\frac{\alpha^R_\mu  \pm \alpha^L_\mu }{2}$ 
transform under the HLS  as 
$\alpha_{\mu ||} 
\to h  \cdot   \alpha_{\mu ||}  \cdot h^\dagger  
- i \partial_\mu h \cdot h^\dagger$ and
$\alpha_{\mu \perp} \to h \cdot   \alpha_{\mu \perp}  \cdot h^\dagger$, respectively. 
Note that $\alpha_{\mu \perp}$ includes the NGB fields as $\alpha_{\mu \perp}= \frac{1}{F_\pi} \partial_\mu \pi + \cdots$. 
The corresponding wave function $(\phi^R - \phi^L)$ should therefore be the eigenfunction for the zero mode, $\psi_0$: 
$  \phi^R (z)-\phi^L(z)  = \psi_0(z)$.  From this and Eq.(\ref{cons:general}) 
we see that the wave functions $\phi^R$ and $\phi^L$ are not the eigenfunctions but are given as 
$  
\phi^{R,L} (z) = \frac{1}{2} \left[ 
1 + \sum_{n=1}^\infty \psi_{2n-1} (z) \pm \psi_0(z) \right] 
$.

By substituting Eq.(\ref{Amuxz}) with Eq.(\ref{cons:general}) into the action (\ref{5d-action}), 
the five-dimensional theory is now described by the NGB fields along with 
the infinite towers of the vector and axialvector meson fields in four dimensions.

We first naively try to truncate towers of the vector and axialvector meson fields 
simply eliminating $V_\mu^{(n)}$ and $A_\mu^{(n)}$ for  $n > N$. 
Then we find  
\begin{eqnarray} 
 S_5^{\rm truncation} 
 \ni 
 \int dz d^4 x \, K_2(z) \sum_{n=N+1}^\infty
\lambda_{2n-1} \psi_{2n-1}^2(z) {\rm tr}[\alpha_{\mu ||} (x^\mu)  ]^2 
\,, \label{action:naive}
\end{eqnarray}  
which explicitly breaks the chiral symmetry as well as the HLS, 
because $\alpha_{\mu ||} \to 
h \cdot   \alpha_{\mu ||}   \cdot h^\dagger  
- i \partial_\mu h \cdot h^\dagger$. 
Naive truncation of tower of vector meson fields thus forces us to encounter the explicit 
violation of the chiral symmetry.

Now we shall propose a method to truncate towers of  vector and axialvector meson fields 
in a gauge-invariant manner.  
Consider a low-energy effective theory below the mass 
of $n=N+1$ level. 
Such an effective theory can be obtained by integrating out 
mesons with $n \ge N+1$ via the equations of motion. 
 Neglecting terms including the derivatives acting on the heavy fields $V_\mu^{(k)}$ 
and $A_\mu^{(k)}$ with $k > N$, 
the equations of motion for them take the following forms:  
$  V_\mu^{(k)} = \alpha_{\mu ||}$, $A_\mu^{(k)} =  0$  
$(k=N+1, N+2, \cdots , \infty) $. 
Putting these solutions into the action, 
we have, instead of Eq.(\ref{action:naive}),  
\begin{eqnarray} 
 S_5^\textrm{integrate out} 
 \ni 
 \int dz d^4 x \, 
K_2(z) \sum_{n=1}^N
\lambda_{2n-1} \psi_{2n-1}^2(z) {\rm tr}[\alpha_{\mu ||} (x^\mu) - V_\mu^{(n)}(x^\mu)]^2   
\,, 
\end{eqnarray} 
which is certainly gauge-invariant. 
Note also that the gauge-invariance now requires not the constraint in Eq.(\ref{cons:general}) 
but $\phi^R(z) + \phi^L(z) - \sum_{n=1}^N \psi_{2n-1}(z) = 1 $.

Let us now consider 
a low-energy effective model obtained by integrating 
out all the higher vector and axialvector meson fields except  
the lowest vector meson field $V_\mu^{(1)} \equiv V_\mu$. 
Following the gauge-invariant way proposed above,  
the expansion of $A_\mu(x^\mu,z)$ is expressed as 
\begin{equation} 
  A_\mu(x^\mu,z) = \alpha_{\mu \perp}(x^\mu) \psi_0(z)
  + (\hat{\alpha}_{\mu ||}(x^\mu) + V_\mu(x^\mu)  )  
  + \hat{\alpha}_{\mu ||}(x^\mu)  \psi_1(z) 
  \,, \label{generalAmu:expand3}
\end{equation} 
where 
$  \hat{\alpha}_{\mu ||}
  = - V_\mu + \alpha_{\mu ||}$. 
One can further introduce the external  gauge fields by gauging the global chiral  
$U(N)_L \times U(N)_R$ symmetry. (For details, see Ref.~\cite{HMY2}.) 
Then we obtain the low-energy effective model 
including only the lightest HLS field $V_\mu$ 
and the NGB fields $\pi$ described by the HLS-ChPT with ${\cal O}(p^4)$ terms~\cite{Tanabashi:1993sr,Harada:2003jx}: 
The ${\cal O}(p^4)$ terms include the effects from infinite towers of higher KK modes 
and are completely determined as explicitly shown in Ref.~\cite{HMY2}; 
sum rules such as those introduced in Ref.~\cite{Sakai:2005yt} 
are also fully built in the HLS-ChPT Lagrangian. 
Our formulation is thus more practical and useful.

Finally, we once again emphasize that our methodology presented here is 
applicable to any models of HQCD.

\section{Application to Sakai-Sugimoto Model}
\label{sec3}

In this section, we apply 
our methodology to the Sakai-Sugimoto (SS) model~\cite{Sakai:2004cn,Sakai:2005yt}
based on $D8/\bar{D}8/D4$ brane configuration. 
The five-dimensional gauge-invariant portion (so-called the Dirac-Born-Infeld (DBI) part) 
of the low-energy effective action in 
the SS model is given by~\cite{Sakai:2004cn,Sakai:2005yt}  
\begin{eqnarray}
S_{\rm SS}^{\rm DBI} &=& N_c G  
        \int d^4x dz  
         \Bigg( 
               - \frac{1}{2} K^{-1/3}(z) {\rm tr}[ 
                                                   F_{\mu\nu} F^{\mu\nu} 
                                                 ]  
               +      K(z) M_{\rm KK}^2  {\rm tr }[ 
                                                   F_{\mu z} F^{\mu z} 
                                                  ] 
         \Bigg)
\,, \label{SSaction}
\end{eqnarray} 
where $K(z) =1 + z^2 $ is the induced metric of the five-dimensional space-time; 
the overall coupling $G$ is the rescaled 't~Hooft coupling expressed as 
$G =  N_cg_{\rm YM}^2/(108 \pi^3) $ with 
$g_{\rm YM}$ being the gauge coupling of 
the $U(N_c)$ gauge symmetry on the $N_c$ D4-branes~\cite{Sakai:2004cn,Sakai:2005yt}; 
the mass scale $M_{KK}$  is related to 
the scale of the compactification of the $N_c$ D4-branes onto the $S^1$. 
Comparing Eq.(\ref{SSaction}) with Eq.(\ref{5d-action}), 
we read off $ K_1(z)=  G K^{-1/3}(z)$, $K_2(z)= G K(z)$, 
so that we find the equation of motion, 
$ -K^{1/3}(z) \partial_z  \left( K(z) \partial_z \psi_n \right) = \lambda_n \psi_n  $ 
with the eigenvalues $\lambda_n$ and the eigenfunctions $\psi_n$ of the KK modes of 
the five-dimensional gauge field $A_\mu(x^\mu, z)$.

Application to the Chern-Simons term is straightforward that is 
explicitly demonstrated in Ref.~\cite{HMY2}.

As emphasized in the end of the previous section, 
without introducing any sum rules, 
we are able to calculate amplitudes 
straightforwardly from the effective Lagrangian which includes contributions from {\it full} set of  higher KK modes. 
To see it more explicitly, as an example, 
we shall study the pion electromagnetic (EM) form factor $F_V^{\pi^\pm}$ 
at tree-level of the present model. 
$F_V^{\pi^\pm}$ is readily 
constructed from the Lagrangian written in terms of the HLS-ChPT 
presented in Ref.~\cite{HMY2}: 
$  
F_V^{\pi^\pm}(Q^2)|_{\rm HLS} 
  = g_{\gamma \pi\pi}(Q^2) + \frac{g_\rho(Q^2) g_{\rho\pi\pi}(Q^2)}{m_\rho^2 + Q^2} 
$, 
where $Q^2=-p^2$ denotes a momentum-squared in space-like region and~\cite{HMY2}  
$ 
g_{\gamma \pi\pi} (Q^2) 
 = 
 \left(1 - \frac{a}{2} \right) + \frac{a g^2z_6}{4} \frac{Q^2}{m_\rho^2}$,  
$g_\rho (Q^2) 
= 
 \frac{m_\rho^2}{g} \left(1 + g^2 z_3 \frac{Q^2}{m_\rho^2} \right)  
 $, 
$g_{\rho \pi \pi} (Q^2) 
= 
 \frac{1}{2} a g \left( 1 + \frac{g^2z_4}{2} \frac{Q^2}{m_\rho^2} \right) 
 $. 
The applicable momentum range should be restricted to  
$0 \le Q^2 \ll \{m_{\rho'}^2, m_{\rho^{\prime\prime}}^2, \cdots\}$ 
since we have integrated out higher KK modes keeping only the $\rho$ meson. 
Note that our form factor $F_V^{\pi^\pm}(Q^2)|_{\rm HLS}$ automatically 
ensures the EM gauge-invariance, $ F_V^{\pi^\pm}(0)|_{\rm HLS}=1 $: 
One can easily show~\cite{HMY2} that 
if towers of vector and axialvector mesons had naively been truncated as in Eq.(\ref{action:naive}) 
one would have $F_V^{\pi^\pm}(0)|_{\rm truncation} \neq 1$, leading to a violation of the EM gauge-invariance. 
(It turns out~\cite{HMY2} that higher KK modes actually play the crucial role to maintain the EM gauge-invariance.) 
We further rewrite $F_V^{\pi^\pm}(Q^2)|_{\rm HLS}$ as~\footnote{ 
Here $g_\rho \equiv g_\rho(Q^2=-m_\rho^2)$ and $g_{\rho\pi\pi}\equiv g_{\rho\pi\pi}(Q^2=-m_\rho^2)$. } 
\begin{equation} 
  F_V^{\pi^\pm} (Q^2) |_{\rm HLS}
  = 
  \left( 1 - \frac{1}{2} \tilde{a} \right) + \tilde{z} \frac{Q^2}{m_\rho^2} 
  + \frac{g_\rho g_{\rho\pi\pi}}{m_\rho^2 + Q^2} 
\,, \label{FV}
\end{equation} 
where $ \tilde{a} = a \left( 1 - \frac{g^2z_4}{2} - g^2z_3 + \frac{(g^2z_3)(g^2z_4)}{2} \right) 
$ and $ \tilde{z} =  \frac{1}{4} a \left(  g^2 z_6 + (g^2 z_3)(g^2 z_4) \right) 
$~\cite{HMY2} which are expressed in terms of the five-dimensional theory as~\footnote{
$\langle A \rangle \equiv \int dz K^{-1/3}(z) A(z)$. }~\cite{HMY2} 
$   \tilde{a} = 
  \frac{2 g_\rho g_{\rho\pi\pi}}{m_\rho^2} 
=\frac{\pi}{4} \lambda_1 
\frac{\langle \psi_1  \rangle  \langle  \psi_1 (1-\psi_0^2) \rangle}{\langle \psi_1^2 \rangle}
$,  
$ \tilde{z} 
 = \frac{\pi}{8} \lambda_1 \left( 
\frac{\langle \psi_1  \rangle  \langle  \psi_1 (1-\psi_0^2) \rangle}{\langle \psi_1^2 \rangle} 
- \langle 1- \psi_0^2 \rangle \right) $. 
These $\tilde{a}$ and $\tilde{z}$ 
are calculated independently of any inputs to be $\tilde{a} \simeq 2.62$ and $\tilde{z} \simeq 0.08$, 
where we have used $\lambda_1 \simeq 0.669$.

\begin{figure}
\begin{center}
\begin{tabular}{cc}
{
\begin{minipage}{0.5\textwidth}
\begin{flushleft}
\end{flushleft}
\includegraphics[scale=0.6]{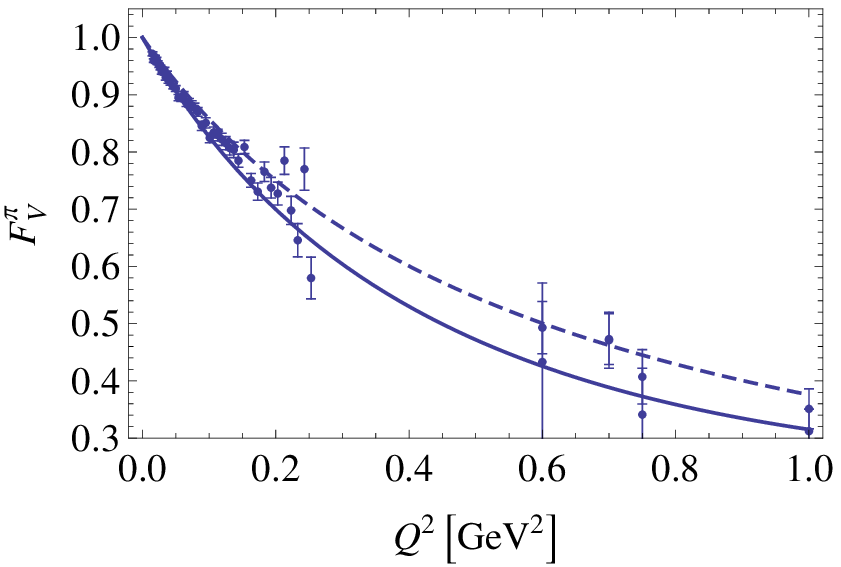} 
\vspace{10pt}  
\end{minipage}
}&
{
\begin{minipage}{0.5\textwidth}
\begin{flushleft} 
\end{flushleft}
\includegraphics[scale=0.6]{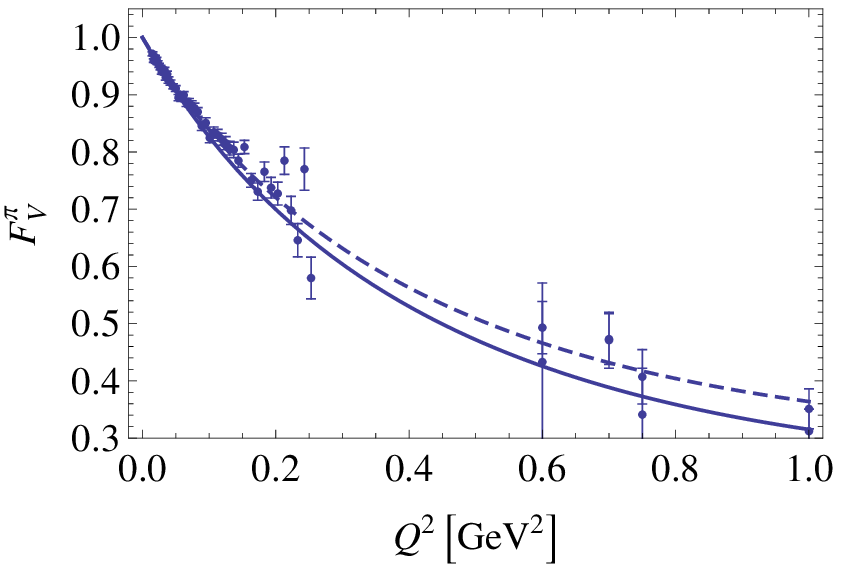} 
\vspace{10pt} 
\end{minipage} 
} 
\end{tabular}
\caption{The predicted curve of the pion EM form factor 
$F_V^{\pi^\pm}$ with respect to space-like momentum-squared $Q^2$ (denoted by solid line) 
fitted with the experimental data from 
Ref.~\cite{Amendolia:1986wj} 
with $\chi^2/{\rm d.o.f}=147/53 \simeq  2.77$. 
In the left panel, 
the dashed curve corresponds to 
the form factor in the LVMD hypothesis of 
the HLS model with ${\tilde a}=2$ and ${\tilde z}=0$ taken. 
($\chi^2/{\rm d.o.f}=226/53 \simeq  4.26$). 
The dashing curve in the right panel 
corresponds to the form factor fitted with the experimental data, 
yielding the best fit values of ${\tilde a}$ and $\tilde{z}$, 
${\tilde a}|_{\rm best}=2.44$, $\tilde{z}|_{\rm best} = 0.08$ 
($\chi^2/{\rm d.o.f}=81/51 \simeq  1.56$). 
}
\label{Fv-holo} 
\end{center}
\end{figure}

Using the expression (\ref{FV}) and the values of $\tilde{a}$ and $\tilde{z}$, 
we evaluate the momentum-dependence of $F_V^{\pi^\pm}$  
which was actually not possible in the original SS model~\cite{Sakai:2005yt} because of  
the form written in terms of the infinite summation. 
In Fig.~\ref{Fv-holo} we show the predicted curve of $F_V^{\pi^\pm}$ 
with respect to $Q^2$ 
together with the experimental data from Ref.~\cite{Amendolia:1986wj}. 
The $\chi^2$-fit results in good agreement with 
the data ($\chi^2/{\rm d.o.f}=147/53 \simeq  2.77$). 
Comparison with the result derived from the lowest vector meson dominance (LVMD)  
hypothesis with ${\tilde a}=2$ and ${\tilde z}=0$ 
is shown by a dashed curve in the left panel of Fig.~\ref{Fv-holo}. 
The right panel of Fig.~\ref{Fv-holo} shows a 
comparison with the result obtained by fitting the parameters $(\tilde{a}, \tilde{z})$ 
to the experimental data. 
It is interesting to note that the best-fit values of $\tilde{a}$ and $\tilde{z}$ 
are quite close to those in the predicted curve. 
This fact reflects  that the predicted curve fits well with the experimental data.

\section{Summary} 
\label{summary}

In this talk, we developed 
a methodology to integrate out 
arbitrary parts of infinite towers of vector and axialvector mesons 
arising as KK modes in a class of HQCD models. 
It was shown that our method is gauge-invariant in contrast to 
a naive truncation [See Eq.(\ref{action:naive})]. 
It was demonstrated that 
any models of HQCD in the low-energy region can be 
described by the HLS-ChPT. 
We applied our method to the SS model and 
evaluated the momentum-dependence of the pion EM form factor 
as an example, which demonstrated power of 
our formulation. 
The predicted form factor was shown to be fitted well with the experimental data 
in the low-energy (space-like momentum) region. 
This was difficult in the original SS model 
due to the forms of the form factors written in terms of infinite sum of vector meson exchanges. 
More on phenomenological applications of our formulation to the SS model is 
presented in Ref.~\cite{HMY2}.

\section*{Acknowledgments} 

This work was supported in part by the JSPS Grant-in-Aid for 
Scientific Research; (B) 18340059 (K.Y.), (C) 20540262 (M.H.), 
Innovative Areas \#2104 
``Quest on New Hadrons with Variety of Flavors'' (M.H.),
the Global COE Program 
``Quest for Fundamental Principles in the Universe''
(K.Y. and M.H.),
and the Daiko Foundation (K.Y.). 
S.M. is supported by the Korea Research 
Foundation Grant funded by the Korean Government (KRF-2008-341-C00008).

\end{document}